# A Cost-Effective Quantum Boolean-Phase SWAP Gate with Only Two CNOT Gates

Ali Al-Bayaty
*Department of Electrical and Computer Engineering*
Portland State University
Portland, OR USA
albayaty@pdx.edu

Shanyan Chen
*College of Information Engineering*
Capital Normal University
Beijing, China
sychen@cnu.edu.cn

Steven A. Bleiler
*Fariborz Maseeh Department of Mathematics and Statistics*
Portland State University
Portland, OR USA
bleilers@pdx.edu

Marek Perkowski
*Department of Electrical and Computer Engineering*
Portland State University
Portland, OR USA
h8mp@pdx.edu

*Abstract*—A new Boolean-Phase swapping gate is presented with improved quantum generality and cost-effectiveness. Our swapping gate is termed the "*p*-SWAP gate", where *p* is the phase difference selected for a set of swapped qubits. The phase *p* is expressed in radians and $-\pi \leq p \leq +\pi$. The generality of *p*-SWAP gate is demonstrated for Phase applications for selected values of *p* and for Boolean applications when the value of *p* is ignored. The cost-effectiveness of *p*-SWAP gate follows from the lower quantum cost and depth of its final realized (transpiled) quantum circuit, when compared to the standard SWAP gate composed of three Feynman (CNOT) gates. Specifically, our presented *p*-SWAP gate utilizes only two CNOT gates. The quantum circuit of the *p*-SWAP gate is visually designed using our previously developed Bloch sphere approach. Experimentally, after transpilation for an IBM quantum computer, the transpiled *p*-SWAP gate shows an approximate 23% quantum cost reduction and an approximate 26% depth minimization compared to the transpiled standard SWAP gate.

## I. Introduction

In quantum computing, the quantum operation of a standard SWAP gate swaps (switches) the basis vectors (as quantum states) of two input qubits into two output qubits [1–3]. These two qubits are termed the "targets". For instance, a standard SWAP gate applies on the targets of $|q_1 q_0\rangle \rightarrow |q_0 q_1\rangle$: $|00\rangle \rightarrow |00\rangle$, $|01\rangle \rightarrow |10\rangle$, $|10\rangle \rightarrow |01\rangle$, and $|11\rangle \rightarrow |11\rangle$, where $q_0$ is the basis vector of the first target and $q_1$ is the basis vector of the second target. In general, the quantum circuit of a standard SWAP gate is constructed using three Feynman (CNOT) gates, where the CNOT gate is considered a cost-expensive 2-bit quantum gate. For IBM quantum computers, the CNOT gate requires multiple microwave pulses more than a 1-bit quantum gate [4, 5]. Notice that the "*n*-bit" is often used instead of the "*n* qubits" for a multiple-qubit gate [6–8], where $n \geq 2$.

The standard SWAP gate is an important design entity in many quantum applications and algorithms. For instance, standard SWAP gates are commonly utilized to: (i) connect the non-neighboring physical qubits of a limited layout (architecture) connectivity for a quantum computer, which is the so-called "heavy-hexagonal" in IBM terminology [9–12], (ii) implement the quantum Fourier Transform (QFT) and the inverse QFT of Shor's algorithm and quantum phase estimation (QPE) algorithm [1–3, 13, 14], (iii) efficiently construct Grover's algorithm and quantum approximate optimization algorithm (QAOA) [15–18], (iv) build quantum Boolean functions and reversible circuits [19–21], (v) implement quantum cryptography and cryptarithmetic algorithms [22–24], (vi) prepare Dicke states [25–27], just to name a few.

For applications considering quantum layouts of a real computer rather than only using simulations with abstract quantum gates, the percentage of generated standard SWAP gates is significantly increased when the number of non-neighboring physical qubits of a computer increases [28, 29]. Subsequently, when a quantum application consisting of many standard SWAP gates is realized (transpiled) into a quantum computer, the transpiled quantum circuit of such an application has a high quantum cost and depth. In this paper, the quantum cost is the total number of the transpiled quantum gates, which are the native "supported" gate of a quantum computer, and the depth is the critical longest path along all native gates. Notice that a higher quantum cost and depth produce a longer delay and increase the decoherence for the utilized physical qubits of a quantum computer [30–32]. For IBM quantum computers of 127 physical qubits [33, 34], their native gates are the Identity (I), Pauli-X (X), half-rotational X ($\sqrt{X}$), CNOT, and rotational Pauli-Z (RZ($\theta$)), where the half-rotation is defined as $+\pi/2$ radians and the rotational is defined as $-\pi \leq \theta \leq +\pi$ radians.

The goal of this paper is to introduce a cost-effective Boolean-Phase swapping gate, which is termed the "*p*-SWAP gate", with a selected phase difference (*p*) for $-\pi \leq p \leq +\pi$ in radians. We designed the *p*-SWAP gate based on two quantum concepts, generality and cost-effectiveness, as follows.

- The generality of the *p*-SWAP gate arises from its utilization in both Boolean and Phase oracles [7, 8, 35] for different quantum applications and algorithms. Such that, the value of *p* can be ignored for the swapped qubits of Boolean oracles or selected in case of Phase oracles.

- The cost-effectiveness of the *p*-SWAP gate arises from its utilization of only two CNOT gates instead of three. Such that, the quantum cost and depth of the *p*-SWAP gate are always lower than those of the standard SWAP gate, after transpilation for a quantum computer.

Our introduced *p*-SWAP gate is derived from our proposed Bloch sphere approach (BSA), and the *p*-SWAP gate is a more generalized phase swapping gate than the *i*SWAP gate proposed in [36] of two CNOT gates. The BSA is a "geometrical design tool" for constructing cost-effective quantum gates, based on their rotational quantum operations in the XY-plane, XZ-plane, YZ-plane, and other parallel planes perpendicular to different coordinate axes with the Bloch sphere. The BSA is discussed in [37] and in our two cost-effective quantum libraries: GALA-*n* (Generic Architecture of Layout-Aware *n*-bit gates) [38, 39] and CALA-*n* (Clifford+T-based Architecture of Layout-Aware *n*-bit gates) [40, 41], where $n \geq 2$ qubits. Both GALA-*n* and CALA-*n* quantum libraries have officially become part of the IBM Qiskit ecosystem [42].

For different quantum computing technologies, various research papers and studies discussed different methods to improve the fidelities, speedups, and circuit depths of the standard SWAP gates, by re-constructing the quantum circuits of the standard SWAP gates using: (i) the momentum-controlled NOT (MC-NOT) and polarization-controlled NOT (PC-NOT) gates for Silicon nanophotonics technology [43], (ii) the $\sqrt{\text{SWAP}}$ gates for Silicon quantum dots technology [44], (iii) the $\sqrt{\text{SWAP}}$ gates for Transmons superconducting technology [45], (iv) the Clifford+T gates for Transmons superconducting technology [46], and (v) the cross-gate pulse cancellation and cross-resonance polarity gates for Transmons superconducting technology [9].

However, for the above-mentioned research papers and studies, we observed that the reconstructed quantum circuits of the standard SWAP gates require at least three CNOT gates. While the quantum circuit of our introduced *p*-SWAP gate only requires two CNOT gates, with a selected value of *p* as well. The purpose of our research is to introduce a generic cost-effective swapping gate with a lower quantum cost and depth for Transmons superconducting technology, e.g., IBM quantum computers.

In this paper, we prove that the *p*-SWAP gate is a generic swapping gate, by ignoring or customizing the value of *p* for a set of swapped qubits. After transpilation using the `ibm_brisbane` quantum computer [33, 34], we experimentally prove that the *p*-SWAP gate is a cost-effective swapping gate for approximately 23% and 26% reduction of its final quantum cost and depth, respectively, as compared to the standard SWAP gate.

## II. METHODS

This section discusses how to visually design the *p*-SWAP gate via the Bloch sphere approach (BSA) in [37], instead of using the conventional design approach of matrix multiplications and unitary representations, to construct a generic cost-effective Boolean-Phase swapping gate consisting of only two CNOT gates.

### A. The Bloch Sphere

The Bloch sphere is a geometrical sphere in a three-dimensional space with the axes X, Y, and Z. This sphere represents the quantum states of a qubit. When a series of quantum gates are applied to a qubit, the Bloch sphere visualizes such quantum operations in Hilbert space ($\mathcal{H}$) [1–3]. Hence, in quantum computing, the Bloch sphere is commonly utilized as a geometrical visualization (and verification) tool. Fig. 1 depicts six distinct states of a qubit, and each visualizes an intersection of one of the three spatial axes with the Bloch sphere.

An *n*-bit gate having X (as a Pauli-X), Y (as a Pauli-Y), or Z (as a Pauli-Z) in its notation rotates the states of a qubit around the X, Y, or Z axis within the Bloch sphere, respectively, where $n \geq 1$. Such a quantum rotation around a spatial axis relies on a defined rotational angle ($\pm \theta$) expressed in radians, as shown in Fig. 1, where $+\theta$ represents a counterclockwise rotation and $-\theta$ represents a clockwise rotation for $-\pi \leq \theta \leq +\pi$. For instance, the following Z gates rotate the states of a qubit by the Z-axis for the defined θ angles:

- The Z gate performs the quantum rotation of $+\pi$ radians.
- The $\sqrt[2]{Z}$ gate performs the quantum rotation of $+\pi/2$ radians. The $\sqrt[2]{Z}$ gate is the so-called "S gate" [1–3, 6].
- The $\sqrt[4]{Z}$ gate performs the quantum rotation of $+\pi/4$ radians. The $\sqrt[4]{Z}$ gate is the so-called "T gate" [1–3, 47].

Notice that all *n*-bit quantum gates are unitary gates [1–3, 6], and a few of them are non-Hermitian gates [1–3, 6], i.e., their quantum operations are not equal to their own inverses, for example, the S and T gates stated above. The S† gate is the inverse gate for the S gate, and the T† gate is the inverse gate for the T gate. In addition, some rotational gates alter the global phase for certain states of a qubit. Thus, the choice of rotational gates can be a critical factor in the circuit design. For instance, the Z gate is not the same as the IBM native RZ($+\pi$) gate, due to the global phase difference of $-i$ between the two gates; such that, $RZ(+\pi) = e^{-i\frac{\pi}{2}} Z = \begin{bmatrix} e^{-i\frac{\pi}{2}} & 0 \\ 0 & e^{i\frac{\pi}{2}} \end{bmatrix} = -iZ$.

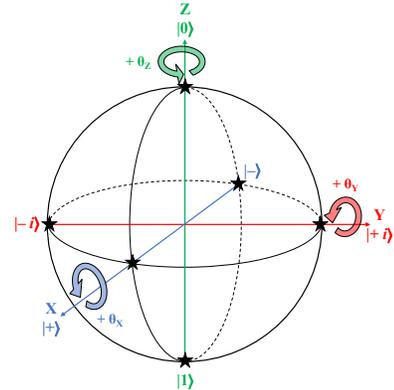

Fig. 1. The Bloch sphere is a three-dimensional space with three axes (X, Y, and Z) for illustrating six distinct states (the stars) of a qubit: (i) the $|0\rangle$ and $|1\rangle$ states intersecting the Z-axis with the Bloch sphere, (ii) the $|+\rangle$ and $|-\rangle$ states intersecting the X-axis with the Bloch sphere, and (iii) the $|+i\rangle$ and $|-i\rangle$ states intersecting the Y-axis with the Bloch sphere. Each axis has its own rotational angle ($\pm \theta$) expressed in radians, where $+\theta$ represents a counterclockwise rotation and $-\theta$ represents a clockwise rotation for $-\pi \leq \theta \leq +\pi$ [37].

## B. Coordinate and Parallel Planar Intersections with The Bloch Sphere

In [37–42], we discussed how to geometrically design any single-target quantum gate using the BSA, using the Bloch sphere and its coordinate and parallel planar intersections. The coordinate and parallel planes are two-dimensional planes perpendicular to the individual three-coordinate axes. In general, the intersections of different coordinate and parallel planes perpendicular to different coordinate axes with the Bloch sphere can be utilized to geometrically design a quantum gate, depending on (i) the quantum operational purpose of such a gate and (ii) the supported native gates of a utilized quantum computer. Hence, the BSA is a "geometrical design tool" that can visually construct any quantum gate. Fig. 2 demonstrates the XY-plane, XZ-plane, YZ-plane, and $m$ parallel planes ($P$) perpendicular to various of the three coordinate axes with the Bloch sphere, where $m \geq 1$ and the black dots denote the segments of "semicircles", "quadrants", and "octants" for the resulting circular intersections with the Bloch sphere.

For IBM quantum computers, because the quantum operations of all native gates mainly rotate the states of a qubit around the X-axis and Z-axis of the Bloch sphere, we utilize the XY-plane of the Bloch sphere (see Fig. 2(a)) to visually design our generic cost-effective $p$-SWAP gate. The XY-plane's intersection with the Bloch sphere is divided into a number of segments, to represent the quantum rotations ($\pm \theta$) of IBM native RZ gates applied to a qubit as follows. We choose a "base" state on each circle of intersection and use the images of that state under various quantum rotations to subdivide the circle of intersection in various ways.

- The "semicircle" segment is half of the XY-plane's intersection with the Bloch sphere that represents the quantum rotations of Z gates, i.e., we use our base state's images under RZ($\pm \pi$) to subdivide our intersection circle. Here then, we consider our intersection circle to consist of two semicircles.

- The "quadrant" segments are obtained by replacing the RZ($\pm \pi$) gate in 1, with either of the S and S[†] gates, i.e., RZ($+ \pi/2$) and RZ($- \pi/2$) and their repetition of gates respectively. Here then, our intersection circle consists of four quadrants.

- Similarly, the "octant" segments are obtained by replacing the RZ($\pm \pi$) gate in 1, with either of the T and T[†] gates, i.e., RZ($+ \pi/4$) and RZ($- \pi/4$) and their repetition of gates respectively. Here then, our intersection circle consists of eight octants.

In the XY-plane, the IBM native X and $\sqrt{X}$ gates rotate the states of a qubit around the X-axis by $+ \pi$ and $+ \pi/2$ radians, respectively. However, the IBM native CNOT gate rotates the state of its target qubit around the X-axis by $+ \pi$ radians, when its control qubit is set to the $|1\rangle$ state; otherwise, there is no rotation occurs. Notice that, for all IBM quantum computers of 127 qubits, the native CNOT gate has been replaced with the native ECR (echoed cross-resonance) gate [48–50] to: (i) minimize the recalibration error rates and coherent errors, and (ii) extend the duration of measurements.

## C. Circuit Design of the p-SWAP Gate

Our generic cost-effective $p$-SWAP gate is the generalization concept of the $i$SWAP gate proposed by the IBM quantum system [36], which only utilizes two CNOT gates. For this, the quantum circuit of the $p$-SWAP gate has only two CNOT gates, as compared to the standard SWAP gate of three CNOT gates shown in Fig. 3(a). Therefore, the $p$-SWAP gate is considered a cost-effective swapping gate in the context of the utilized number of CNOT gates.

The $i$SWAP gate, shown in Fig. 3(b), swaps the states of its two targets as a standard SWAP gate. However, the $i$SWAP gate only alters the phases to $i$ (a phase difference of $+ \pi/2$ radians), when the two targets are initially in the $|01\rangle$ and $|10\rangle$ states. In contrast, our $p$-SWAP gate can select different values of the phase difference ($p$) for a set of the targets in the $|00\rangle$, $|01\rangle$, $|10\rangle$, and/or $|11\rangle$ states, where $-\pi \leq p \leq +\pi$ in radians. Therefore, the $p$-SWAP gate is considered a generic swapping gate in the context of different values selection of $p$.

On the one hand, the $p$-SWAP gate is a generic cost-effective Phase swapping gate for Phase oracles, since this gate is customized to the value of $p$. On the other hand, when the value of $p$ is ignored, i.e., $p$ can be arbitrarily set to any angular value, such a $p$-SWAP gate is a generic cost-effective Boolean swapping gate for Boolean oracles. Subsequently, the $p$-SWAP gate is a generic cost-effective Boolean-Phase swapping gate.

For IBM quantum computers, the circuit of the $p$-SWAP gate is entirely constructed using IBM native $\sqrt{X}$, RZ, and CNOT gates. Such that, its transpiled quantum circuit always has a lower quantum cost and depth, as compared to the $i$SWAP gate having IBM non-native Hadamard (H) gates. Notice that each H gate is decomposed into a sequence {RZ($+ \pi/2$), $\sqrt{X}$, RZ($+ \pi/2$)} of IBM native gates after transpilation, which yields to increase the quantum cost and depth for the transpiled quantum circuit of the $i$SWAP gate. For this, in the circuit design of the $p$-SWAP gate, we utilized the technique of replacing all H gates with the $\sqrt{X}$ gates to create uniform superposition states of all qubits. Such a technique for performing a cost-effective transpilation is discussed in [51].

The quantum circuit of the $p$-SWAP gate is visually designed via the BSA, using the Bloch sphere and the XY-plane, instead of using the conventional design via unitary representations and multiplications. Such a visual design mainly relies on the following four quantum operations, as "stages" as depicted in Fig. 3(c). Notice that initial states for the two targets of the $p$-SWAP gates are on the Z-axis intersected with the Bloch sphere.

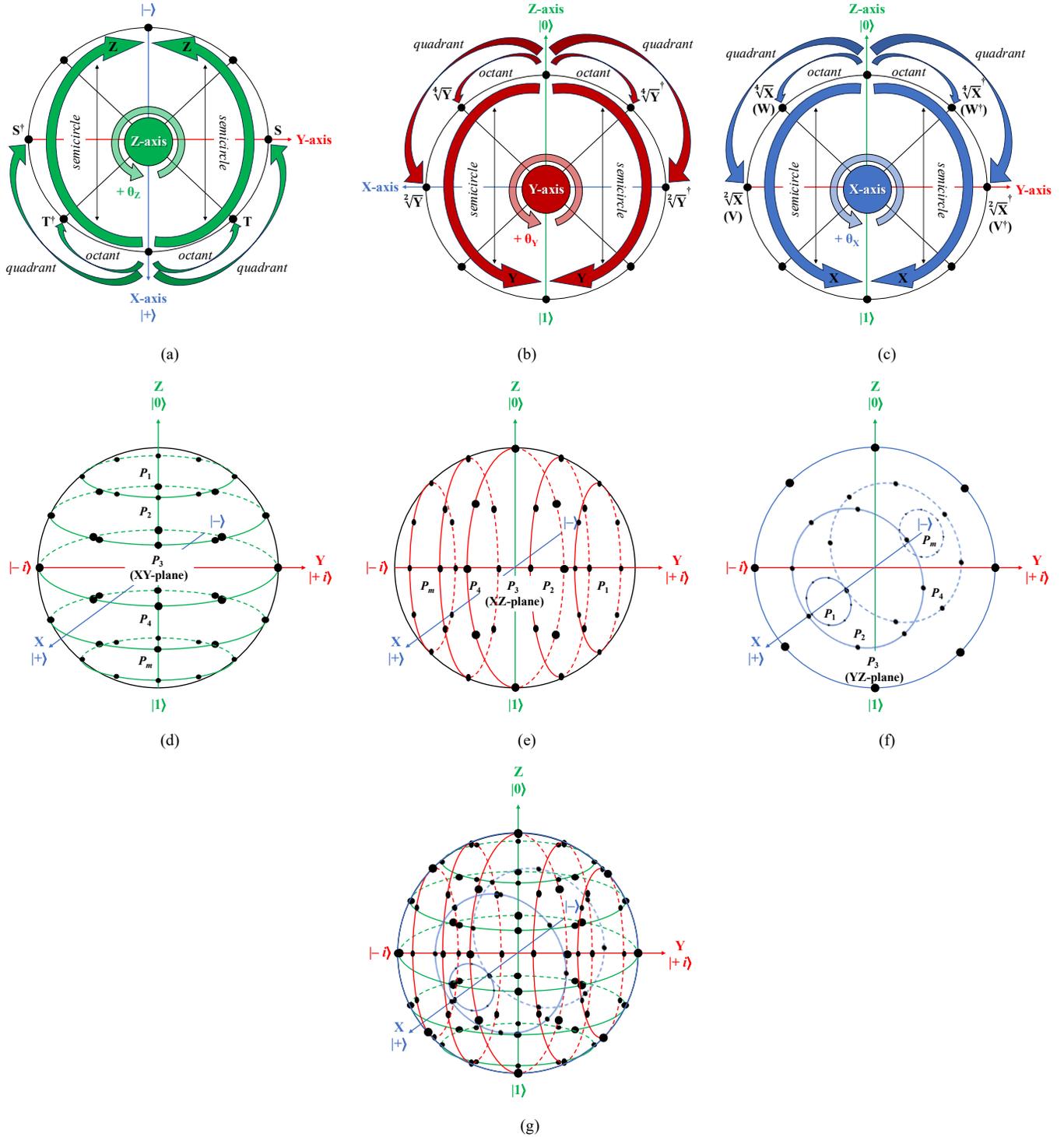

Fig. 2. Coordinate and parallel planes (*P*) perpendicular to various coordinate axes intersected with the Bloch sphere: (a) the XY-plane perpendicular to the Z-axis intersected with the Bloch sphere (as $P_3$ in d), (b) the XZ-plane perpendicular to the Y-axis intersected with the Bloch sphere(as $P_3$ in e), (c) the YZ-plane perpendicular to the X-axis intersected with the Bloch sphere (as $P_3$ in f), (d) *m* parallel planes perpendicular to the Z-axis, (e) *m* parallel planes perpendicular to the Y-axis, (f) *m* perpendicular to the X-axis, and (g) 3 collections of *m* parallel planes, each collection perpendicular to one of the three coordinate axes of the Bloch sphere, where $m \geq 1$ and the black dots subdividing their intersections with the Bloch sphere into the semicircles, quadrants, and octants for each such intersections [37].

- *Entanglement stage*: creates a uniform superposition state for the first target using the $\sqrt{X}$ gate, i.e., places the first target on the Y-axis of the XY-plane. The second target is then entangled with the superposition of states for the first target using the CNOT gate.

- *Differentiation stage*: differentiates how the entangled states of the two targets are correlated via rotating them by choosing appropriate angular values, to replace such entangled states with entangled states on the other side of the Y-axis in the XY-plane.

- *Unentanglement stage*: collapses the resulting entangled states of the first target using the CNOT gate controlled by the states of the second target. The first target is thus placed back on the Z-axis. The unentangled states of the second target are then placed back on the Z-axis using the $\sqrt{X}$ gate.

- *Phase selection stage*: selects the values of $p$ for a set of the swapped targets using two 1-bit phase-based gates, which are denoted as $v$ and $w$ gates for the first and second targets, respectively.

Notice that, for these stages, all utilized quantum gates are IBM native gates, including the 1-bit phase-based gates ($v$ and $w$), which are RZ gates. However, for a different quantum computer, the IBM native gates of the $p$-SWAP gate can be replaced with the supported native gates of such a quantum computer. In such a case, a different parallel plane ($P$) of the BSA (see Fig. 2) should be carefully considered when visually designing a new cost-effective $p$-SWAP gate. Therefore, in this paper, we introduced the BSA as an open visual circuit design framework and the $p$-SWAP gate as a generic cost-effective Boolean-Phase swapping gate for any quantum computer.

As illustrated in Fig. 3(c), for IBM quantum computers, the quantum circuit of the $p$-SWAP gate for Boolean oracles is constructed utilizing the first three stages without the phase selection stage. In contrast, the quantum circuit of the $p$-SWAP gate for Phase oracles is constructed utilizing all four stages. Notice that: (i) the BSA is used here to geometrically choose appropriate angular values for the RZ($\theta_1$) and RZ($\theta_2$) gates of the differentiation stage, and (ii) the chosen angular values for these two RZ gates must fulfill the requirement of the unentanglement stage.

Since the $p$-SWAP gate is a two-target quantum gate, two XY-planes (one plane for each target) are required to visually design such a gate, by geometrically choosing the appropriate angular values of the RZ gates of the differentiation stage. Based on these stages, Fig. 4 demonstrates the quantum transitions for the swapped outcomes of a $p$-SWAP gate (see Figure 3(c)) when applied to all permutative states ($q_0$ and $q_1$) of the two targets. For ease of illustration, we abusively represent the entangled states of the two targets on two Bloch spheres. Practically, two Bloch spheres could not visualize the maximally entangled states of two qubits [52–54]. However, the maximally entangled states of two qubits are easily represented in the toric geometry map of the joint state-space of two qubits [55], which is not widely known in the quantum engineering community. From Fig. 4, the angular values for both RZ gates in the differentiation stage are geometrically chosen for $\theta_1 = \theta_2 = +\pi/2$ radians, i.e., each RZ($+\pi/2$) gate moves a state through one quadrant of the XY-plane intersected the Bloch sphere (see Fig. 2(a)). This fulfills the requirement of the unentanglement stage. While the two 1-bit phase-based gates ($v$ and $w$) of the phase selection stage remain undetermined, they will be discussed in the Results section.

Thereby, in this paper, we illustrated how to utilize the BSA to visually design the $p$-SWAP gates as generic cost-effective Boolean-Phase swapping gates, fulfilling the quantum demands of Boolean and Phase oracles arising in specific-task quantum applications and algorithms.

### III. RESULTS

For quantum phase-based applications, different 1-bit phase-based gates ($v$ and $w$ shown in Fig. 3(c)) are chosen to obtain particular values of $p$ for the swapped outcomes of a $p$-SWAP gate, as expressed in Table I, where None means a gate does not exist in the phase selection stage and $-\pi \leq p \leq +\pi$ in radians. On the one hand, in Table I, specific angular values for RZ gates ($v$ and $w$) are chosen for prospective quantum phase-based applications as follows.

- Voting for a set of the swapped outcomes, based on their particular values of $p$.
- Constructing a cost-effective $i$SWAP gate.
- Constructing a cost-effective $i$SWAP inverse gate.

On the other hand, different angular values for RZ gates ($v$ and $w$) can be assigned and evaluated for other values of $p$, since the phase selection stage of a $p$-SWAP gate is considered an interesting topic for developing and investigating various quantum phase-based applications.

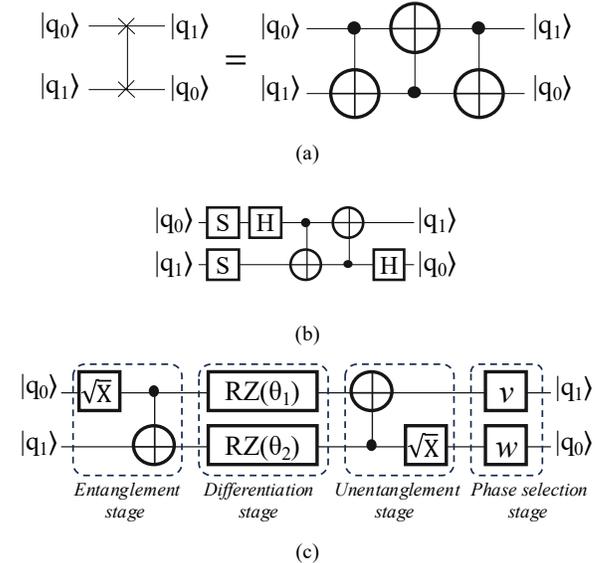

Fig. 3. Decomposed quantum circuits of SWAP, $i$SWAP, and $p$-SWAP gates: (a) the standard SWAP gate, (b) the $i$SWAP gate [36], and (c) the $p$-SWAP gate consisting of IBM native gates only, four stages, and two 1-bit phase-based gates ($v$ and $w$) for selecting the values of $p$, where $|q_0\rangle$ is the initial state of the first target, $|q_1\rangle$ is the initial state of the second target, and $-\pi \leq p \leq +\pi$ in radians.

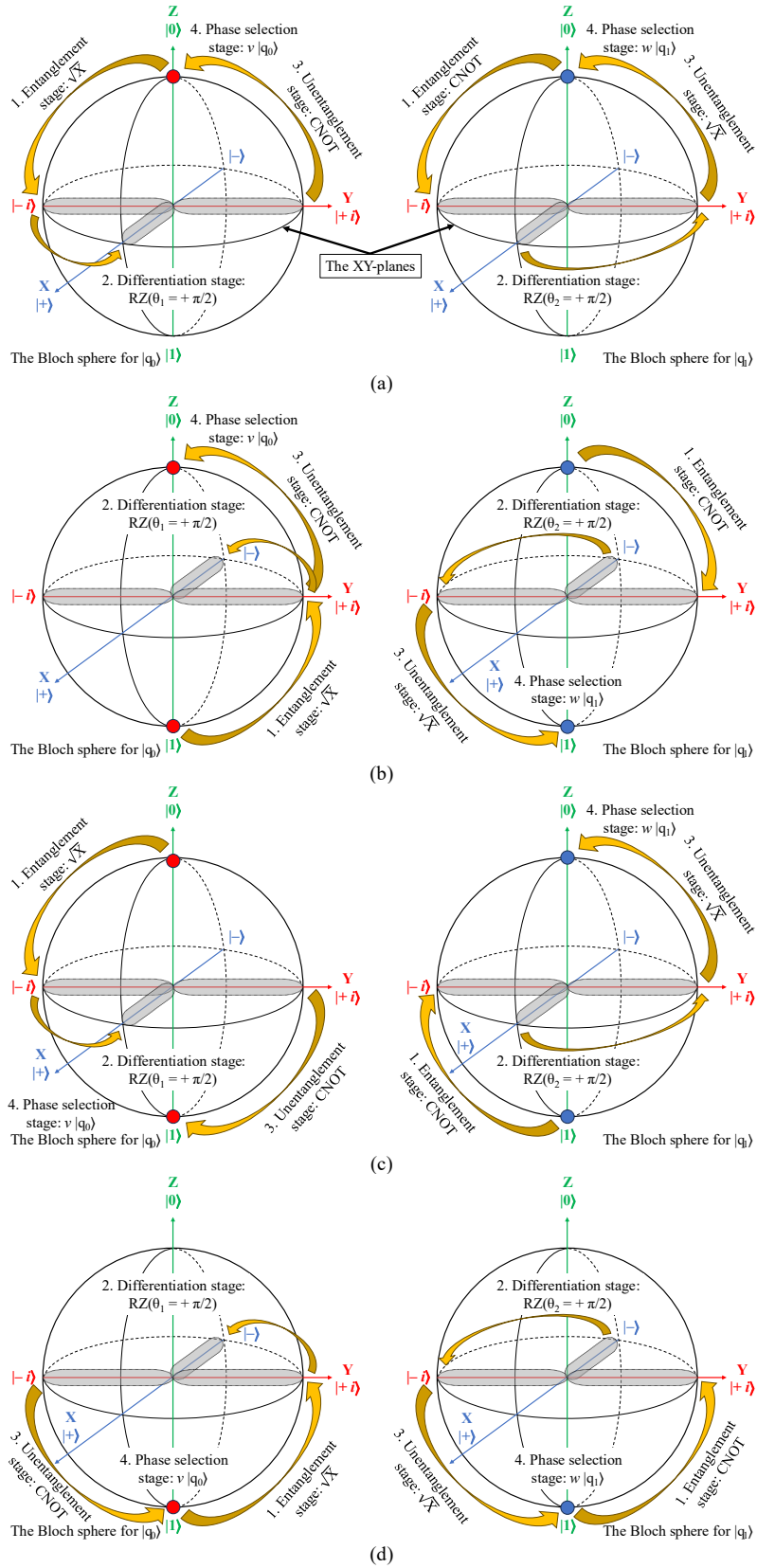

Fig. 4. Representations of the quantum transitions (the yellow arrows) for the swapped outcomes of a *p*-SWAP gate using two XY-planes intersecting two Bloch spheres, for all permutative states $|q_1q_0\rangle$: (a) $|00\rangle \rightarrow |00\rangle$, (b) $|01\rangle \rightarrow |10\rangle$, (c) $|10\rangle \rightarrow |01\rangle$, and (d) $|11\rangle \rightarrow |11\rangle$, where the red dots represent the states of the first target, the blue dots represent the states of the second target, and the shaded gray areas abusively represent the entangled states of the two targets.

TABLE I. APPLICATIONS FOR THE SWAPPED OUTCOMES OF A *P*-SWAP GATE USING PARTICULAR VALUES OF *P*

| RZ gates | | Phase *p* for the swapped outcomes | | | | Phase-based applications |
|---|---|---|---|---|---|---|
| *v* | *w* | $\|00\rangle$ | $\|01\rangle$ | $\|10\rangle$ | $\|11\rangle$ | |
| None | None | 0 | 0 | $+\pi$ | 0 | Voting for the state $\|10\rangle$ only |
| RZ($+\pi$) | None | $-\pi/2$ | $+\pi/2$ | $+\pi/2$ | $+\pi/2$ | Voting for the state $\|00\rangle$ only |
| None | RZ($+\pi$) | $-\pi/2$ | $-\pi/2$ | $-\pi/2$ | $+\pi/2$ | Voting for the state $\|11\rangle$ only |
| RZ($+\pi$) | RZ($-\pi$) | 0 | $+\pi$ | 0 | 0 | Voting for the state $\|01\rangle$ only |
| RZ($+\pi/2$) | RZ($-\pi/2$) | 0 | $+\pi/2$ | $+\pi/2$ | 0 | The *i*SWAP gate, voting for both states $\|01\rangle$ and $\|10\rangle$ |
| RZ($-\pi/2$) | RZ($+\pi/2$) | 0 | $-\pi/2$ | $-\pi/2$ | 0 | The *i*SWAP inverse gate, voting for both states $\|01\rangle$ and $\|10\rangle$ |

In our research, the quantum circuits of the standard SWAP gate (shown in Fig. 3(a)), the *i*SWAP gate (shown in Fig. 3(b)), and our *p*-SWAP gate (shown in Fig. 3(c)) are transpiled and evaluated using the `ibm_brisbane` quantum computer of 127 qubits [33, 34]. These transpiled quantum circuits are evaluated in the context of our proposed "transpilation quantum cost (TQC)" [38, 56]. The TQC is the sum of all 1-bit IBM native $\sqrt{X}$, X, and RZ gates (denoted as $N_1$), all 2-bit IBM native ECR gates (denoted as $N_2$), and the depth (denoted as D), where D is the critical longest path through all $N_1$ and $N_2$ gates. Fig. 5(a) depicts the transpilation of one CNOT gate into an ECR gate using the `ibm_brisbane` quantum computer. Notice that the technical specifications [28–30, 57–59] of this quantum computer, e.g., the relaxation time (T1), decoherence time (T2), mean readout error of native gates, circuit layer operations per second (CLOPS), and quantum volume (QV), are not considered for the purpose of our research.

After transpilation, it was concluded that the transpiled quantum circuit of the *p*-SWAP gate has lower TQC and depth than those of the transpiled quantum circuit of the standard SWAP gate, for approximately 23% and 26% reduction of TQC and depth, respectively. However, the transpiled quantum circuit of the *p*-SWAP gate has identical TQC and depth to those of the transpiled quantum circuit of the *i*SWAP gate, due to the fact that the *p*-SWAP gate is the generalization concept of the *i*SWAP gate. Table II summarizes the calculated TQC for the transpiled quantum circuits of the standard SWAP, *i*SWAP, and *p*-SWAP gates, as shown in Fig. 5(b), Fig. 5(c), and Fig. 5(d), respectively.

Therefore, the *p*-SWAP gate can be employed as a generic and cost-effective transpilation library for the IBM quantum system, to replace the conventional transpilation of the standard SWAP gates, with the customizability of *p* based on the quantum operational purpose of the utilized Boolean and Phase oracles in various quantum applications and algorithms.

## IV. CONCLUSION

The standard quantum SWAP gate switches the basis vectors (quantum states) of its two input qubits into two output qubits, which are termed the "targets". This standard SWAP gate is commonly constructed using three Feynman (CNOT) gates. Thus, this gate is considered a cost-expensive quantum gate, when it is realized (transpiled) into a quantum computer. In this paper, we design a generic cost-effective swapping gate constructed using only two CNOT gates. We term our swapping gate as the "*p*-SWAP gate", where *p* is a customizable phase difference for $-\pi \leq p \leq +\pi$ in radians. We introduce the *p*-SWAP gate as a generic cost-effective swapping framework. The gate's generality arises from the customizability of *p* for a set of swapped targets for Phase applications, and for Boolean applications when *p* is ignored. The gate's cost-effectiveness arises from the fewer utilized number of CNOT gates, as compared to the standard SWAP gate.

The quantum circuit of the *p*-SWAP gate is derived from the quantum circuit of the *i*SWAP gate proposed by the IBM quantum system in [36], and the customizability of *p* is geometrically chosen using our proposed Bloch sphere approach (BSA). Notice that when the value of *p* is set to $+\pi/2$ radians for the states $|01\rangle$ and $|10\rangle$ only, then the *p*-SWAP gate behaves identically to the *i*SWAP gate. In general, the BSA is a "geometrical design tool" for constructing cost-effective quantum gates, based on their rotational quantum operations on a plane perpendicular to one of the three coordinate axes intersected with the Bloch sphere, without using the conventional design approach of matrix multiplications and unitary representations.

Experimentally, the quantum circuits of the standard SWAP and our *p*-SWAP gates were transpiled and evaluated on an IBM quantum computer. We conclude that the transpiled quantum circuit of the *p*-SWAP gate always has a lower quantum cost and depth than those of the transpiled quantum circuit of the standard SWAP gate. In conclusion, the *p*-SWAP gate can be employed as a generic and cost-effective transpilation design framework, which can replace the conventional framework for the standard SWAP gates, with the customizability of *p* fulfilling the quantum demands of Boolean and Phase applications.

TABLE II. SUMMARY OF TQC FOR THE TRANSPILED QUANTUM CIRCUITS OF THE STANDARD SWAP, *I*SWAP, AND *P*-SWAP GATES

| | $N_1$ | $N_2$ | D | TQC |
|---|---|---|---|---|
| Standard SWAP gate: | 20 | 3 | 15 | **38** |
| IBM *i*SWAP gate: | 16 | 2 | 11 | **29** |
| Our *p*-SWAP gate: | 16 | 2 | 11 | **29** |

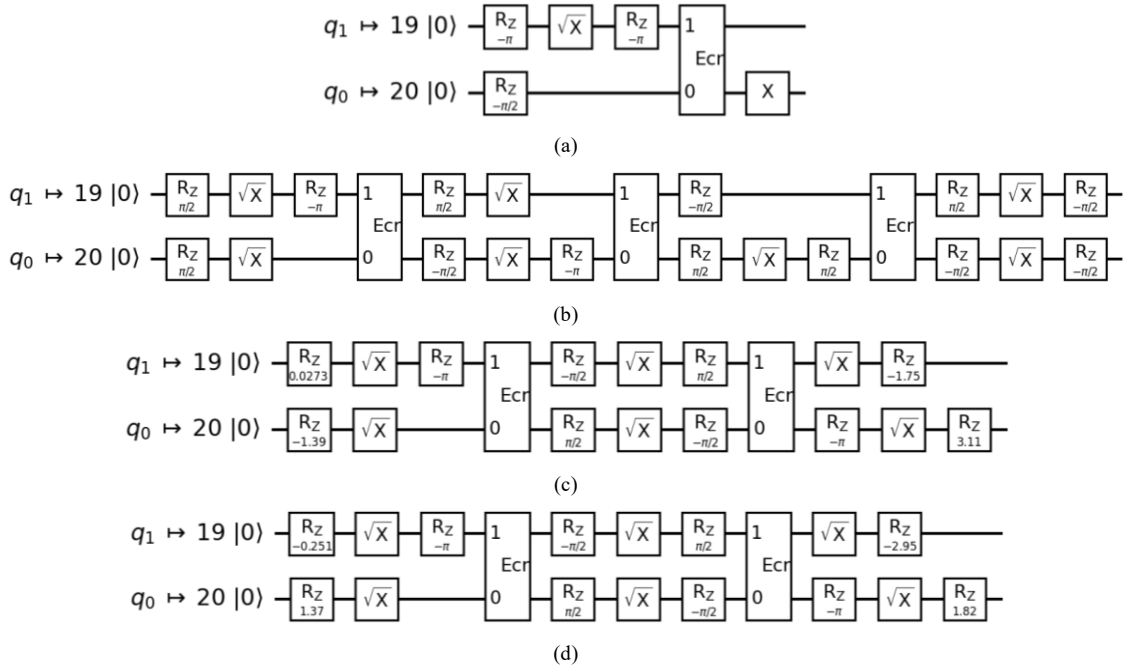

Fig. 5. Transpiled quantum circuits using `ibm_brisbane` quantum computer: (a) one Feynman (CNOT) gate, (b) the standard SWAP gate consisting of three CNOT gates, (c) the *i*SWAP gate consisting of two CNOT gates, and (d) our generic cost-effective *p*-SWAP gate consisting of two CNOT gates. Note that, for all IBM quantum computers of 127 qubits, the native CNOT gate has been replaced with the native ECR (echoed cross-resonance) gate incorporated with the native $\sqrt{X}$, X, and RZ gates [48–50].

## REFERENCES

[1] P. Kaye, R. Laflamme, and M. Mosca, *An Introduction to Quantum Computing*. New York, NY, USA: Oxford University Press, 2007.

[2] M.A. Nielsen and I.L. Chuang, *Quantum Computation and Quantum Information*, 10th ed. Cambridge, UK: Cambridge University Press, 2010.

[3] R. LaPierre, *Introduction to Quantum Computing*. Cham, Switzerland: Springer, 2021.

[4] N. Earnest, C. Tornow, and D.J. Egger, "Pulse-efficient circuit transpilation for quantum applications on cross-resonance-based hardware," *Physical Review Research*, vol. 3, no. 4, p. 043088, 2021.

[5] T. Hurant and D.D. Stancil. "Asymmetry of CNOT gate operation in superconducting Transmon quantum processors using cross-resonance entangling," 2020, *arXiv:2009.01333*.

[6] A. Barenco, C.H. Bennett, R. Cleve, D.P DiVincenzo, N. Margolus, P. Shor, T. Sleator, J.A. Smolin, and H. Weinfurter, "Elementary gates for quantum computation," *Physical Review A*, vol. 52, no. 5, p. 3457, 1995.

[7] A. Al-Bayaty and M. Perkowski, "A concept of controlling Grover diffusion operator: a new approach to solve arbitrary Boolean-based problems," *Scientific Reports*, vol. 14, p. 23570, 2024.

[8] A. Al-Bayaty and M. Perkowski, "BHT-QAOA: The generalization of quantum approximate optimization algorithm to solve arbitrary Boolean problems as Hamiltonians," *Entropy*, vol. 26, no. 10, p. 843, 2024.

[9] P. Gokhale, T. Tomesh, M. Suchara, and F. Chong, "Faster and more reliable quantum swaps via native gates," in *Proc. of the 2024 Int. Conf. on Parallel Architectures and Compilation Techniques*, 2024, pp. 351–362.

[10] S.F. Lin, S. Sussman, C. Duckering, P.S. Mundada, J.M. Baker, R.S. Kumar, A.A. Houck, and F.T. Chong, "Let each quantum bit choose its basis gates," in *2022 55th IEEE/ACM Int. Symp. on Microarchitecture (MICRO)*, 2022, pp. 1042–1058.

[11] C. Duckering, J.M. Baker, A. Litteken, and F.T. Chong, "Orchestrated trios: compiling for efficient communication in quantum programs with 3-qubit gates," in *Proc, of the 26th ACM Int. Conf. on Architectural Support for Programming Languages and Operating Systems*, 2021, pp. 375–385.

[12] V. Gheorghiu, J. Huang, S.M. Li, M. Mosca, and P. Mukhopadhyay, "Reducing the CNOT count for Clifford+T circuits on NISQ architectures," *IEEE Trans. on Computer-Aided Design of Integrated Circuits and Systems*, vol. 42, no. 6, pp.1873–1884, 2022.

[13] A. Adedoyin, J. Ambrosiano, P. Anisimov, W. Casper, G. Chennupati, C. Coffrin, H. Djidjev, D. Gunter, S. Karra, N. Lemons, and S. Lin, "Quantum algorithm implementations for beginners,", 2018, *arXiv:1804.03719*.

[14] D. Camps, R. Van Beeumen, and C. Yang, "Quantum Fourier transform revisited," *Numerical Linear Algebra with Applications*, vol. 28, no. 1, p. e2331, 2021.

[15] A. Mandviwalla, K. Ohshiro, and B. Ji, "Implementing Grover's algorithm on the IBM quantum computers," in *2018 IEEE Int. Conf. on Big Data*, 2018, pp. 2531–2537.

[16] J. Weidenfeller, L.C. Valor, J. Gacon, C. Tornow, L. Bello, S. Woerner, and D.J. Egger, "Scaling of the quantum approximate optimization algorithm on superconducting qubit based hardware," *Quantum*, vol. 6, p. 870, 2022.

[17] K. Blekos, D. Brand, A. Ceschini, C.H. Chou, R.H. Li, K. Pandya, and A. Summer, "A review on quantum approximate optimization algorithm and its variants," *Physics Reports*, vol. 1068, pp. 1–66, 2024.

[18] B. Tan and J. Cong, "Optimal layout synthesis for quantum computing," in *Proc. of the 39th Int. Conf. on Computer-Aided Design*, 2020, pp. 1–9.

[19] Y. Guowu, W.N.N. Hung, and S. Xiaoyu, "Majority-based reversible logic gates," *Theoretical Computer Science*, vol. 334, no. 13, p. 259274, 2005.

[20] M. Lukac, S. Nursultan, G. Krylov, and O. Keszöcze, "Geometric refactoring of quantum and reversible circuits: quantum layout," in *2020 23rd Euromicro Conf. on Digital System Design (DSD)*, 2020, pp. 428–435.

[21] M. Lukac, M. Perkowski, H. Goi, M. Pivtoraiko, C.H. Yu, K. Chung, H. Jeech, B.G. Kim, and Y.D. Kim, "Evolutionary approach to quantum and reversible circuits synthesis," *Artificial Intelligence Review*, vol. 20, pp. 361–417, 2003.

[22] K. Jang, G. Song, H. Kim, H. Kwon, H. Kim, and H. Seo, "Efficient implementation of PRESENT and GIFT on quantum computers," *Applied Sciences*, vol. 11, no. 11, p. 4776, 2021.

[23] S. Jaques, M. Naehrig, M. Roetteler, and F. Virdia, "Implementing Grover oracles for quantum key search on AES and LowMC," in *Advances in Cryptology–EUROCRYPT 2020: 39th Ann. Int. Conf. on the Theory and Applications of Cryptographic Techniques*, 2020, pp. 280–310.

[24] P. Fernández and M.A. Martin-Delgado, "Implementing the Grover algorithm in homomorphic encryption schemes," *Physical Review Research*, vol. 6, no. 4, p. 043109, 2024.

[25] A. Bärtschi and S. Eidenbenz, "Deterministic preparation of Dicke states," in *Int. Symp. on Fundamentals of Computation Theory*, 2019, pp. 126–139.

[26] C.S. Mukherjee, S. Maitra, V. Gaurav, and D. Roy, "Preparing Dicke states on a quantum computer," *IEEE Trans. on Quantum Engineering*, vol. 1, pp. 1–17, 2020.

[27] S. Aktar, A. Bärtschi, A.H.A. Badawy, and S. Eidenbenz, "A divide-and-conquer approach to Dicke state preparation," *IEEE Trans. on Quantum Engineering*, vol. 3, pp. 1–16, 2022.

[28] V. Pina-Canelles, A. Auer, and I. de Vega, "Improving and benchmarking NISQ qubit routers," 2025, *arXiv:2502.03908*.

[29] J. Liu, P. Li, and H. Zhou, "Not all swaps have the same cost: a case for optimization-aware qubit routing," in *2022 IEEE Int. Symp. on High-Performance Computer Architecture (HPCA)*, 2022, pp. 709–725.

[30] A. Mi, S. Deng, and J. Szefer, "Securing reset operations in NISQ quantum computers," in *Proc. of the 2022 ACM SIGSAC Conf. on Computer and Communications Security*, 2022, pp. 2279–2293.

[31] P. Murali, J.M. Baker, A. Javadi-Abhari, F.T. Chong, and M. Martonosi, "Noise-adaptive compiler mappings for noisy intermediate-scale quantum computers," in *Proc. of the 24th Int. Conf. on Architectural Support for Programming Languages and Operating Systems*, 2019, pp. 1015–1029.

[32] D. Koch, B. Martin, S. Patel, L. Wessing, and P.M. Alsing, "Demonstrating NISQ era challenges in algorithm design on IBM's 20 qubit quantum computer," *AIP Advances*, vol. 10, no. 9, 2020.

[33] R.C. Farrell, M. Illa, A.N. Ciavarella, and M.J. Savage, "Scalable circuits for preparing ground states on digital quantum computers: the Schwinger model vacuum on 100 qubits," *PRX Quantum*, vol. 5, no. 2, p. 020315, 2024.

[34] IBM Quantum Platform. "Quantum Processing Units." IBM.com. Accessed: Mar. 14, 2025. [Online]. Available: https://quantum.ibm.com/services/resources?tab=systems&system=ibm_brisbane

[35] C. Figgatt, D. Maslov, K.A. Landsman, N.M. Linke, S. Debnath, and C. Monroe, "Complete 3-qubit Grover search on a programmable quantum computer," *Nature Communications*, vol. 8, no. 1, p.1918, 2017.

[36] IBM Quantum Documentation. "iSwapGate." IBM.com. Accessed: Mar. 14, 2025. [Online]. Available: https://docs.quantum.ibm.com/api/qiskit/qiskit.circuit.library.iSwapGate

[37] A. Al-Bayaty and M. Perkowski, "BSA: the Bloch sphere approach as a geometrical design tool for building cost-effective quantum gates," *Protocols.io*, 2024, doi: 10.17504/protocols.io.bp2l6dkkdvqe/v4.

[38] A. Al-Bayaty and M. Perkowski, "GALA-n: generic architecture of layout-aware n-bit quantum operators for cost-effective realization on IBM quantum computers," 2023, *arXiv:2311.06760*.

[39] A. Al-Bayaty. "GALA-n Quantum Library." GITHUB.com. Accessed: Mar. 14, 2025. [Online]. Available: https://github.com/albayaty/gala_quantum_library

[40] A. Al-Bayaty, X. Song, and M. Perkowski, "CALA-n: a quantum library for realizing cost-effective 2-, 3-, 4-, and 5-bit gates on IBM quantum computers using Bloch sphere approach, Clifford+T gates, and layouts," 2024, *arXiv:2408.01025*.

[41] A. Al-Bayaty. "CALA-n Quantum Library." GITHUB.com. Accessed: Mar. 14, 2025. [Online]. Available: https://github.com/albayaty/cala_quantum_library

[42] IBM Quantum Computing. "Qiskit Ecosystem." IBM.com. Accessed: Mar. 14, 2025. [Online]. Available: https://www.ibm.com/quantum/ecosystem

[43] X. Cheng, K.C. Chang, Z. Xie, M.C. Sarihan, Y.S. Lee, Y. Li, X. Xu, A.K. Vinod, S. Kocaman, M. Yu, and P.G.O. Lo, "A chip-scale polarization-spatial-momentum quantum SWAP gate in Silicon nanophotonics," *Nature Photonics*, vol. 17, no. 8, pp. 656–665, 2023.

[44] M. Ni, R.L. Ma, Z.Z. Kong, X. Xue, S.K. Zhu, C. Wang, A.R. Li, N. Chu, W.Z. Liao, G. Cao, and G.L. Wang, "A SWAP gate for spin qubits in Silicon," 2023, *arXiv:2310.06700*.

[45] J. Howard, A. Lidiak, C. Jameson, B. Basyildiz, K. Clark, T. Zhao, M. Bal, J. Long, D.P. Pappas, M. Singh, and Z. Gong, "Implementing two-qubit gates at the quantum speed limit," *Physical Review Research*, vol. 5, no.4, p. 043194, 2023.

[46] P. Niemann, L. Mueller, and R. Drechsler, "Combining SWAPs and remote CNOT gates for quantum circuit transformation," in *2021 24th Euromicro Conf. on Digital System Design (DSD)*, 2021, pp. 495–501.

[47] B. Schmitt and G. De Micheli, "Tweedledum: a compiler companion for quantum computing," in *Design, Automation & Test in Europe Conf. & Exhibition (DATE)*, 2022, pp. 7–12.

[48] P. Jurcevic, A. Javadi-Abhari, L.S. Bishop, I. Lauer, D.F. Bogorin, M. Brink, L. Capelluto, O. Günlük, T. Itoko, N. Kanazawa, and A. Kandala, "Demonstration of quantum volume 64 on a superconducting quantum computing system," *Quantum Science and Technology*, vol. 6, no. 2, p. 025020, 2021.

[49] E. Pelofske, A. Bärtschi, and S. Eidenbenz, "Quantum volume in practice: what users can expect from NISQ devices," *IEEE Trans. on Quantum Engineering*, vol. 3, pp. 1–19, 2022.

[50] Y. Ji, K.F. Koenig, and I. Polian, "Optimizing quantum algorithms on Bipotent architectures," *Physical Review A*, vol. 108, no. 2, p. 022610, 2023.

[51] A. Al-Bayaty and M. Perkowski, "Superposition states on different axes of the Bloch sphere for cost-effective circuits realization on IBM quantum computers," 2023, *arXiv:2311.09326*.

[52] F. Verstraete, K. Audenaert, and B. De Moor, "Maximally entangled mixed states of two qubits," *Physical Review A*, vol. 64, no. 1, p. 012316, 2001.

[53] O. Gamel, "Entangled Bloch spheres: Bloch matrix and two-qubit state space," *Physical Review A*, vol. 93, no. 6, p. 062320, 2016.

[54] C.R. Wie, "Two-qubit Bloch sphere," *Physics*, vol. 2, no. 3, pp. 383–396, 2020.

[55] I. Bengtsson, J. Brännlund, and K. Zyczkowski, "CP^n, or, entanglement illustrated," *Int. Journal of Modern Physics A*, vol. 17, no. 31, pp. 4675–4695, 2002.

[56] A. Al-Bayaty and M. Perkowski, "Cost-effective realization of n-bit Toffoli gates for IBM quantum computers using the Bloch sphere approach and IBM native gates," 2024, *arXiv:2410.13104*.

[57] D.C. McKay, I. Hincks, E.J. Pritchett, M. Carroll, L.C. Govia, and S.T. Merkel, "Benchmarking quantum processor performance at scale," 2023, *arXiv:2311.05933*.

[58] A. Wack, H. Paik, A. Javadi-Abhari, P. Jurcevic, I. Faro, J.M. Gambetta, and B.R. Johnson, "Quality, speed, and scale: three key attributes to measure the performance of near-term quantum computers," 2021, *arXiv:2110.14108*.

[59] A.W. Cross, L.S. Bishop, S. Sheldon, P.D. Nation, and J.M. Gambetta, "Validating quantum computers using randomized model circuits," *Physical Review A*, vol. 100, no. 3, p. 032328, 2019.